\documentclass[12pt,a4paper]{iopart}

\usepackage{iopams}  
\usepackage{xspace}
\usepackage{graphicx}
\usepackage{dsfont}
\usepackage{color}

\newtheorem{theorem}{Theorem}[section] 
\newtheorem{definition}[theorem]{Definition}

\newtheorem{remark}[theorem]{Remark}

\usepackage[bookmarks=true,
            bookmarksopen=true,colorlinks=false]{hyperref}

\newcommand{\R}{\mathds R\xspace}

\newcommand{\So}{\ensuremath{\mathbb S^1}\xspace}

\newcommand{\Sth}{\ensuremath{\mathbb S^3}\xspace}
\newcommand{\eqref}[1]{\eref{#1}}
\newcommand{\ee}{\mathrm e}
\newcommand{\ii}{\mathrm i}
\newcommand{\dd}{\mathrm d}
\newcommand{\E}{\mathcal E}
\newcommand{\Al}{\mathcal A_1}
\newcommand{\Ar}{\mathcal A_2}
\newcommand{\Hp}{{\mathcal H_\mathrm{p}}}
\newcommand{\Hf}{{\mathcal H_\mathrm{f}}}
\newcommand{\Hps}{{\mathcal H_\mathrm{p}'}}
\newcommand{\Hfs}{{\mathcal H_\mathrm{f}'}}
\newcommand{\p}{_\mathrm{p}}
\newcommand{\f}{_\mathrm{f}}

\newcommand{\KVa}{\partial_{\rho_1}}
\newcommand{\KVb}{\partial_{\rho_2}}
\newcommand{\Sp}{{\mathcal S_\mathrm{p}}}
\newcommand{\Sf}{{\mathcal S_\mathrm{f}}}

\begin{document}

\title[Gowdy-symmetric cosmological models with Cauchy horizons ruled by \dots]{Gowdy-symmetric cosmological models with Cauchy horizons ruled by non-closed null generators}

\author{J\"org Hennig}
\address{Department of Mathematics and Statistics,
University of Otago, P.O. Box 56, Dunedin 9054, New Zealand}
\eads{\mailto{jhennig@maths.otago.ac.nz}}

\begin{abstract}
Smooth Gowdy-symmetric generalized Taub-NUT solutions are a class of inhomogeneous cosmological models with spatial three-sphere topology. They have a past Cauchy horizon with closed null-generators, and they have been shown to  to develop a second, regular Cauchy horizon in the future, unless in special, well-defined singular cases. Here we generalize these models to allow for past Cauchy horizons ruled by \emph{non-closed} null generators. In particular, we show local and global existence of such a class of solutions with two functional degrees of freedom. This removes a periodicity condition for the asymptotic data at the past Cauchy horizon that was required before. Moreover, we derive a three-parametric family of exact solutions within that class and study its properties.
\end{abstract}


\section{Introduction}

Gowdy-symmetric solutions to Einstein's field equations are useful test cases for studying key issues in cosmology, e.g.\ questions related to strong cosmic censorship, spikes and the BKL conjecture. Moreover, their investigation promises insights that clearly go beyond the realm of the homogeneous standard models. Consequently, these spacetimes are of considerable interest --- especially in mathematical general relativity.

A particular class of Gowdy-symmetric models has been introduced in a recent paper \cite{BeyerHennig2012}: the \emph{smooth Gowdy-symmetric generalized Taub-NUT (SGGTN) solutions}. These spacetimes are generally not known in the form of exact solutions. (Nevertheless, an example of a three-parametric family of exact solutions has been provided in \cite{BeyerHennig2014}.) Instead, existence has been shown with abstract methods. The main result was that for any choice of two free functions, subject to a periodicity condition, a unique SGGTN solution exists. Each SGGTN spacetime has a past Cauchy horizon, and the two free functions describe the behaviour of the spacetime geometry as this horizon is approached. Moreover, with the exception of particular ``singular cases'', which will be explained below, these solutions develop a regular second, future Cauchy horizon.

The goal of the present article is to investigate the nature of the initial constraint (the periodicity condition), which restricts the choice of data at the past horizon. We will see that this condition is related to a topological feature of this horizon, namely the fact that the null generators have closed orbits. Here, we extend the construction of SGGTN spacetimes to situations with horizons ruled by non-closed generators. The resulting class of spacetimes turns out to again have two functional degrees of freedom, but a periodicity condition is not required anymore. Note that many investigations of cosmological Cauchy horizons in the literature assume closedness of the generators, mainly for technical reasons (see, e.g., \cite{FriedrichRaczWald1999,MoncriefIsenberg1983,Racz2000}). Therefore, it is particularly interesting to study properties of spacetimes for which this assumption is violated.

The following abstract investigation of SGGTN solutions with a more general past horizon will be complemented by the construction of a family of exact solutions. This enables us to study some properties of such spacetimes in more detail. In particular, we construct and investigate extensions of the exact solutions through the Cauchy horizons. 

This paper is organized as follows. In Sec.~\ref{sec:SGGTN}, we give an overview of the construction of SGGTN solutions, which summarizes key steps in the local and global existence proofs presented in \cite{BeyerHennig2012}. Afterwards, in Sec.~\ref{sec:general}, we generalize these models to incorporate Cauchy horizons with non-closed null generators. The announced family of exact solutions within this class is then derived and studied in Sec.~\ref{sec:exact}. Finally, in Sec.~\ref{sec:discuss}, we discuss our findings.

\section{Smooth Gowdy-symmetric generalized Taub-NUT solutions\label{sec:SGGTN}}
We start by summarizing some properties of the class of smooth Gowdy-symmetric generalized Taub-NUT (SGGTN) solutions, before we generalize these models in Sec.~\ref{sec:general}. Details can be found in \cite{BeyerHennig2012}. 

The SGGTN cosmological models are Gowdy-symmetric vacuum solutions to Einstein's field equations with spatial three-sphere topology $\Sth$. It is always possible to achieve the following form of the metric in terms of a time coordinate $t$ and coordinates $\theta$, $\rho_1$ and $\rho_2$ on the three-spheres,
\begin{equation}\label{eq:metric}
  g_{ab}=\ee^M(-\dd t^2+\dd\theta^2)+R_0\left[\sin^2\!t\,\ee^u (\dd\rho_1+Q \dd\rho_2)^2+\sin^2\!\theta\,\ee^{-u} \dd\rho_2^2\right].
\end{equation}
Here, $R_0$ is a positive constant and $u$, $Q$ and $M$ are functions of $t$ and $\theta$ alone. The spatial coordinates are defined in the regions
\begin{equation}\label{eq:domains}
 \theta\in(0,\pi),\quad
 \frac{\rho_1+\rho_2}{2}\in(0,2\pi),\quad 
 \frac{\rho_1-\rho_2}{2}\in(0,2\pi).
\end{equation}
In the following we will outline how existence of globally hyperbolic solutions was shown in the time interval $t\in(0,\pi)$. Also note that an important property of the SGGTN models is their extendibility beyond this domain as non-globally hyperbolic spacetimes, very similar to the famous Taub solution \cite{Taub1951} and its extensions to Taub-NUT \cite{Misner1963, MisnerTaub1969, ChruscielIsenberg1993}.  

In our coordinates, the two Killing vectors corresponding to Gowdy-symmetry can be chosen to be $\KVa$ and $\KVb$. The norm of $\KVa$ is $R_0\sin^2\!\theta\,\ee^u$, which vanishes at $t=0$, provided $u$ is bounded there. Indeed, the SGGTN solutions have a past Cauchy horizon at $t=0$, which we denote as $\Hp$. This horizon is a smooth null surface with $\Sth$-topology and its null generator is proportional to $\KVa$. Moreover, the orbits of this null generator are closed curves.

Note that, here and in the following we will, for the sake of simplicity, follow our conventions from \cite{BeyerHennig2012, BeyerHennig2014} and consider quantities ``at $t=0$'' or ``at $t=\pi$'', even though, strictly speaking, our coordinates break down there. But what we actually mean, of course, are the boundaries that formally correspond to $t=0,\pi$ in  suitable regular coordinates. Similarly, we will often simply call certain Killing vectors ``generators'' of a Cauchy horizon, even though the actual generators, i.e.\ the tangent vectors to the corresponding null geodesics, are only proportional to, but not identical with these vectors.

With the previous observations, and in view of the local and global existence results below, we can state precisely what exactly we mean with the term ``SGGTN solutions'':
\begin{definition}[Initial definition of SGGTN solutions]\label{def1}
 SGGTN solutions are the class of all Gowdy-symmetric vacuum solutions with spatial topology $\Sth$ for which, in terms of the coordinates $t$, $\theta$, $\rho_1$, $\rho_2$ with line element \eqref{eq:metric}, the metric functions $u(t,\theta)$ and $Q(t,\theta)$ at $t=0$ are smooth functions on $\Sth$ subject to the periodicity condition $u(0,0)=u(0,\pi)$ in a Killing basis for which $Q$ has the boundary values $Q(0,0)=1$, $Q(0,\pi)=-1$.
\end{definition}
\begin{remark}\mbox{}

 \begin{enumerate}
  \item The local and global existence results show that this class of solutions is not empty and, more precisely, that it has two functional degrees of freedom.
  \item It follows from this definition that SGGTN solutions always have a smooth past Cauchy horizon at $t=0$ that is generated by the Killing vector $\partial_{\rho_1}$ and has closed orbits. On the other hand, it is not guaranteed that every $\Sth$ Gowdy-symmetric vacuum solution with a smooth past Cauchy horizon ruled by closed null-generators is contained in this class. The reason is that, while the coordinates above can always be chosen to describe a Gowdy-symmetric solution with $\Sth$-topology in a time interval that is contained in $(0,\pi)$, the coordinates become singular at $t=0$. For regular metric potentials, this singular behaviour enforces existence of a smooth Cauchy horizon, but it cannot be excluded that there are solutions for which $u$ and $Q$ are not well-behaved at $t=0$ --- just because of the coordinate singularity there, but which still give rise to a smooth solution with a regular past Cauchy horizon.
 \item The main goal of this paper is to enlarge the class of SGGTN solutions by removing the periodicity condition for $u$ (or, equivalently for the function $S_{**}$ below), which will allow the past horizon to have non-closed null generators. For that purpose, we will later replace this definition with Def.~\ref{def2} below.
 \end{enumerate}
\end{remark}

The Einstein vacuum equations for the metric \eqref{eq:metric} lead to two second order equations for $u$ and $Q$, which are independent of $M$,
\begin{equation}\label{eq:u}
 -u_{,tt}-\cot t\,u_{,t}+u_{,\theta\theta}
 +\cot\theta\,u_{,\theta}
 +\ee^{2u}\frac{\sin^2\! t}{\sin^2\!\theta}\left[Q_{,t}^{\ 2}
 -Q_{,\theta}^{\ 2}\right] +2 = 0,
\end{equation}
\begin{equation}\label{eq:Q}
 -Q_{,tt}-3\cot t\,Q_{,t}+Q_{,\theta\theta}
 -\cot\theta\,Q_{,\theta}-2(u_{,t}Q_{,t} - u_{,\theta}Q_{,\theta})=0
\end{equation}
(as usual commas denote partial derivatives)
and two first-order equations for $M$,
\begin{eqnarray}\fl\label{eq:M1}
 (\cos^2\! t-\cos^2\!\theta)M_{,t} & = &
  \frac{1}{2}\ee^{2u}\frac{\sin^3\! t}{\sin\theta}
 \Big[\cos t\sin\theta\left[Q_{,t}^{\ 2}+Q_{,\theta}^{\ 2}\right]
       -2\sin t\cos\theta\, Q_{,t}Q_{,\theta}\Big]
 \nonumber\\
 & & +\frac{1}{2}\sin t \sin\theta
 \Big[\cos t\sin\theta\left[u_{,t}^{\ 2}+u_{,\theta}^{\ 2}\right]
       -2\sin t\cos\theta\,u_{,t}u_{,\theta}\Big]
 \nonumber\\
 & & -(2\cos^2\!t\,\cos^2\!\theta\,-\cos^2\!t-\cos^2\!\theta)
      \,u_{,t}
 \nonumber\\
 & & -2\sin t\cos t\sin\theta\cos\theta(u_{,\theta}+\tan\theta),  
\end{eqnarray}
\begin{eqnarray}\label{eq:M2}\fl
 (\cos^2\! t-\cos^2\!\theta)M_{,\theta} & = &
  -\frac{1}{2}\ee^{2u}\frac{\sin^3\! t}{\sin\theta}
 \Big[\sin t\cos\theta\left[Q_{,t}^{\ 2}+Q_{,\theta}^{\ 2}\right]
       -2\cos t\sin\theta\,Q_{,t}Q_{,\theta}\Big]
 \nonumber\\
 & & -\frac{1}{2}\sin t \sin\theta
 \Big[\sin t\cos\theta\left[u_{,t}^{\ 2}+u_{,\theta}^{\ 2}\right]
       -2\cos t\sin\theta\, u_{,t}u_{,\theta}\Big]
 \nonumber\\ 
 & & -2\sin t\cos t\sin\theta\cos\theta(u_{,t}-\tan t)
 \nonumber\\
 & & -(2\cos^2\!t\,\cos^2\!\theta\,-\cos^2\!t-\cos^2\!\theta)
      \,u_{,\theta}.
\end{eqnarray}  
In addition, there is a second-order equation for $M$, which, however, is equivalent to the integrability condition $M_{,t\theta}=M_{,\theta t}$ of the first-order system. 

The metric potentials must have a particular behaviour near the axes $\Al$ ($\theta=0$) and $\Ar$ ($\theta=\pi$) as a consequence of the spatial $\Sth$-topology. For our choice of coordinates and Killing basis we obtain, in particular, boundary conditions for $Q$, namely
\begin{equation}\label{eq:Qcond}
 \Al:\quad Q=Q_1=+1,\quad
 \Ar:\quad Q=Q_2=-1,
\end{equation}
and for $M$,
\begin{equation}\label{eq:Mcond}
 \mathcal A_{1/2}:\quad \ee^{M+u}=R_0.
\end{equation}

Note that it was by no means clear that the Einstein equations admit smooth solutions $u$, $Q$, $M$  with the required axes behaviour, i.e.\ it was not guaranteed that there are any SGGTN solutions (apart from the \emph{analytic} solutions studied by Moncrief \cite{Moncrief1984}). However, existence --- first near the past Cauchy horizon and then globally for $t\in(0,\pi$) --- can be shown as follows. In order to prove local existence near $t=0$, the Fuchsian methods developed in \cite{Ames2013a, Ames2013b} have been used in \cite{BeyerHennig2012} to study a singular initial value problem for the Einstein equations with asymptotic data at $t=0$. In a first step, one can show that smooth solutions $u$ and $Q$ exist in some neighbourhood of $t=0$. Afterwards it was shown that the first-order equations for $M$ are then integrable, provided the asymptotic data satisfy a periodicity condition. More precisely, the following local existence result has been shown.
\begin{theorem}\label{Thm1}
 Let $S_{**}$ and $Q_{*}$ be axially symmetric functions
  in $C^\infty(\mathbb S^2)$ subject to the periodicity condition
  \begin{equation}\label{eq:percon}
   S_{**}(0)=S_{**}(\pi)
  \end{equation}
  and $R_0$ a positive constant.  Then there exists a unique smooth
  Gowdy-symmetric generalized Taub-NUT solution for all $t\in(0,\delta]$ (for a sufficiently small $\delta>0$) 
  satisfying the following uniform expansions at $t=0$:
  \begin{eqnarray}
    \label{eq:uexpan}
    R_0\, \ee^{u(t,\theta)} &=& \ee^{S_{**}(\theta)}+\mathcal O(t^2),\\
    \label{eq:Qexpan}
    Q(t,\theta)&=&\cos\theta+Q_*(\theta)\sin^2\!\theta 
    +\mathcal O(t^2),\\
    M(t,\theta)&=&S_{**}(\theta)-2S_{**}(0)+2\ln R_0+\mathcal O(t^2).
  \end{eqnarray}
\end{theorem}
This theorem also shows that the SGGTN models have two degrees of freedom, namely the two smooth functions $S_{**}(\theta)$ and $Q_{*}(\theta)$, subject to the periodicity condition \eqref{eq:percon}.

Once local existence was ensured, global existence in the time interval $0<t<\pi$ followed from a result due to Chru\'sciel \cite{Chrusciel1990}.

Note that the line element \eqref{eq:metric} degenerates at $t=0$ and at $t=\pi$. The degeneracy at $t=0$ comes from a coordinate singularity at the Cauchy horizon (which can be removed by introducing suitable regular coordinates). However, since the global existence result does only apply for $t<\pi$, the question about the behaviour at $t=\pi$, where the metric again degenerates, remained.

In order to study the situation at $t=\pi$, it is useful to reformulate the two Einstein equations \eref{eq:u}, \eref{eq:Q} for $u$ and $Q$  in form of the Ernst equation
\begin{equation}\label{eq:EE}
 \Re(\E) \left(-\E_{,tt}-\cot t\,\E_{,t}
        +\E_{,\theta\theta}+\cot\theta\,\E_{,\theta}\right)
 =-\E_{,t}^{\ 2}+\E_{,\theta}^{\ 2}
\end{equation}
for the complex Ernst potential $\E=f+\ii b$, where $f=\Re(\E)$ is defined by
\begin{equation}\label{eq:deff}
 f:=\frac{1}{R_0}g(\partial_{\rho_2},\partial_{\rho_2})
 =Q^2\ee^u\sin^2\! t+\ee^{-u}\sin^2\!\theta
\end{equation}
and $b=\Im(\E)$ is given by
\begin{equation}\label{eq:defb}
 a_{,t}=\frac{1}{f^2}\sin t\sin\theta\,b_{,\theta},\quad
 a_{,\theta}=\frac{1}{f^2}\sin t\sin\theta\,b_{,t}
\end{equation}
with
\begin{equation}\label{eq:defa}
 a:= \frac{g(\partial_{\rho_1},\partial_{\rho_2})}
          {g(\partial_{\rho_2},\partial_{\rho_2})}
   = \frac{Q}{f}\ee^u\sin^2 t.
\end{equation}
Interestingly, the Ernst equation \eqref{eq:EE} belongs to the remarkable class of integrable equations, i.e.\ there is an associated linear matrix problem which is equivalent to the original nonlinear equation via its integrability condition. Consequently, methods from soliton theory can be applied. In particular, it is possible to integrate the linear problem along the boundaries $t=0,\pi$ and $\theta=0,\pi$, which allows one to find explicit formulae for the Ernst potential at the axes $\Al$, $\Ar$ and  at $t=\pi$ in terms of the data at $\Hp$ ($t=0$). From $\E$ one can then construct the corresponding metric potentials $u$, $Q$ and $M$ at these boundaries, so that the behaviour of the metric at $t=\pi$ can be studied. 
From the explicit expressions for the metric at $t=\pi$ on can just read off how the solution behaves at this boundary, with the following result.

 SGGTN solutions, which are characterized by a past Cauchy horizon $\Hp$ at $t=0$, develop a second, future Cauchy $\Hf$ at $t=\pi$. 
 The only exceptions are special cases in which curvature singularities form. This occurs when the initial data satisfy
 \begin{equation}
  b_B-b_A= \pm4,
 \end{equation}
 where $b_A=b(t=0,\theta=0)$ and $b_B=b(t=0,\theta=\pi)$. In these cases, the solutions have a curvature singularity at $t=\pi$, $\theta=0$ (for a `$+\!$' sign) or at $t=\pi$, $\theta=\pi$ (for a `$-\!$' sign), respectively. If the data at $t=0$ satisfy $b_A=b_B$, then the future horizon is generated by $\KVa$ (as is the past horizon). For $b_A\neq b_B$, $\Hf$ is generated by the linear combination $\KVa-a\f\KVb$, where
 \begin{equation}\label{eq:af0}
  a\f=\frac{8(b_B-b_A)}{16+(b_B-b_A)^2}.
 \end{equation}
 In this case, the metric potential $u$ blows up in the limit $t\to\pi$ even though the spacetime is regular there.

The construction of SGGTN solutions described above assumes from the beginning that the past horizon $\Hp$ is generated by $\KVa$. Due to the periodicity of the coordinates $\rho_1$ and $\rho_2$, this implies that the orbits of $\KVa$ (and $\KVb$) are closed. In Sec.~\ref{sec:general}, we will allow for past horizons generated by linear combinations of both Killing vectors (but initially still using the same coordinates and the same Killing basis as here). This generally corresponds to generators with non-closed orbits, unless the coefficients in the linear combination are commensurable. In particular, we will see that this makes the periodicity condition \eqref{eq:percon} unnecessary. Instead, the asymptotic data function $S_{**}$ can then take on different values at $\theta=0$ and $\theta=\pi$ on $\Hp$, where the size of the ``jump'' in $S_{**}$ is related to the coefficients in the linear combination. 

\section{Generalization to past horizons with non-closed null generators\label{sec:general}}

As indicated above, we will now generalize the SGGTN solutions to situations where the past horizon $\Hp$ is generated by 
$\partial_{\rho_1}-a\p\partial_{\rho_2}$, $a\p=\textrm{constant}$. We will see shortly that the coefficient $a\p$ is the constant value of the Ernst auxiliary quantity $a$ [defined in \eqref{eq:defa}] at $\Hp$, which explains our notation $a\p$. The earlier SGGTN models with a past horizon generated by $\KVa$ are contained in our generalization as the special case $a\p=0$.

\subsection{Setup}

As in the case $a\p=0$, we start from the metric \eqref{eq:metric}. Moreover, 
as a consequence of the $\Sth$-topology, $Q$ and $M$ again satisfy the boundary conditions \eqref{eq:Qcond} and \eqref{eq:Mcond} at the axes.

For $a\p=0$, we observed \cite{BeyerHennig2012} that $u$ is regular at $\Hp$ (which was generated by $\partial_{\rho_1}$), whereas $\ee^u$ diverged as $1/\sin^2\!t$ at $\Hf$ (which was generated by a linear combination of $\KVa$ and $\KVb$) --- unless $b_A=b_B$, in which case $u$ was also regular at $\Hf$ (then generated by $\partial_{\rho_1}$).

Motivated by this observation, we intend to show that for $a\p\neq0$, we find solutions for which
\begin{equation}
 \ee^u=\frac{\ee^v}{\sin^2\!t}
\end{equation}
such that $\ee^v$ is regular at $\Hp$.

In terms of $v$, the metric reads
\begin{equation}\label{eq:metric2}
  g_{ab}=\ee^M(-\dd t^2+\dd\theta^2)+R_0\left[\ee^v (\dd\rho_1+Q \dd\rho_2)^2+\sin^2\! t\,\sin^2\!\theta\,\ee^{-v} \dd\rho_2^2\right].
\end{equation}
It follows from the Einstein equation for $Q$, evaluated in the limit $t\to 0$ and assuming regularity of $\ee^v$, that $Q=Q\p=\textrm{constant}$ at $\Hp$. Furthermore, the definition of the Ernst auxiliary quantity $a$ implies that
\begin{equation}
 a=\frac{g(\partial_{\rho_1},\partial_{\rho_2})}{g(\partial_{\rho_2},\partial_{\rho_2})}
  = \frac{Q\ee^v}{Q^2\ee^v+\ee^{-v}\sin^2\! t\,\sin^2\!\theta}
  \stackrel{t\to 0}{\longrightarrow}
  \frac{1}{Q},
\end{equation}
whence
\begin{equation}\label{eq:aQ}
 a|_{\Hp}\equiv a\p=\frac{1}{Q\p}=\textrm{constant}.
\end{equation}
Then we find
\begin{eqnarray}\fl
 g(\partial_{\rho_1}-a\p\partial_{\rho_2},\partial_{\rho_1}-a\p\partial_{\rho_2})
 &=& g_{\rho_1\rho_1}-2a\p g_{\rho_1\rho_2}+a\p^2 g_{\rho_2\rho_2}\nonumber\\
\fl
 &=& R_0[(1-a\p Q)^2\ee^v+a\p^2\sin^2\! t\,\sin^2\!\theta\,\ee^{-v}]
 \stackrel{t\to 0}{\longrightarrow} 0,
\end{eqnarray}
i.e.\ $\Hp$ is indeed generated by $\partial_{\rho_1}-a\p\partial_{\rho_2}$. In the following we will show that such solutions exist locally and globally, and the future Cauchy horizon $\Hf$ (which exists with the exception of singular cases) is generated by $\partial_{\rho_1}-a\f\partial_{\rho_2}$, where the value of the constant $a\f$ can be computed from the data at $\Hp$.

For our following discussions, we will assume that $a\p\neq \pm1$ (and therefore $Q\p\neq \pm1$) holds. Otherwise (as we will see later), $\Hp$ would be irregular. Indeed, also in the earlier case $a\p=0$, not the past but the future horizon $\Hf$ was regular unless $b_B-b_A=\pm 4$, in which case $a$ takes on the value $\pm 1$ on $\Hf$, cf.~\eqref{eq:af0}. Hence, also in that case the irregularity was related to the horizon boundary value $\pm1$ of $a$. Since our definition of SGGTN solutions assumes a regular past horizon, we can exclude this case.

As a consequence of the boundary values $Q_1=1$, $Q_2=-1$ and $Q\p\neq\pm1$, $Q$ is discontinuous at the points $A$ ($t=\theta=0$) and $B$ ($t=0$, $\theta=\pi$) --- in contrast to the earlier case $a\p=0$, where $Q$ was continuous. Therefore, it follows from the requirement of continuity of the metric coefficients that $\ee^v$ must vanish at $A$ and $B$. Hence we will sometimes denote $N=\ee^v$, such that $N$ is regular at $\Hp$, and $N=0$ at $A$ and $B$. 

First, we will investigate the behaviour of $Q$ and $N$ near the points $A$ and $B$ in more detail. These results will then be used in Subsection~\ref{sec:rotation} to show that a rotation of the Killing basis translates this problem almost exactly into the problem that we had earlier (for $a\p=0$), merely the meaning of the new $\rho$-coordinates will be slightly different and $M$ will satisfy modified boundary conditions at $\Al$ and $\Ar$. Hence we can apply our, almost unchanged, earlier existence results to the new situation with a general past horizon. Only the integrability of the $M$-equations will then require a ``jump'' condition replacing the earlier periodicity condition \eqref{eq:percon}, see Eq.~\eqref{eq:jumpcond} below.
\subsection{The Einstein equations near $A$ and $B$}

We now assume for a moment that SGGTN solutions with a general past horizon exist and derive some properties of the metric potentials for this case, based on the assumption of regularity. Afterwards we will show that solutions with the required properties indeed exist.

The Einstein equations for $N\equiv\ee^v$ and $Q$ are
\begin{eqnarray}
 \fl 0 &=& -N_{,tt}+N_{,\theta\theta}+\frac{N_{,t}^{\ 2}-N_{,\theta}^{\ 2}}{N}
       -\cot t\, N_{,t}+\cot\theta\,N_{,\theta}
       +\frac{N^3}{\sin^2\!t\,\sin^2\!\theta}(Q_{,t}^{\ 2}-Q_{,\theta}^{\ 2}),\\
 \fl 0 &=& -Q_{,tt}+\cot t\,Q_{,t}+Q_{,\theta\theta}-\cot\theta\,Q_{,\theta}
           -\frac{2}{N}(N_{,t}Q_{,t}-N_{,\theta}Q_{,\theta}).
\end{eqnarray}
In order to study the behaviour of $N$ and $Q$ near $A$, we introduce polar coordinates $(r,\alpha)$ centred at $A$,
\begin{equation}
 \theta=r\cos\alpha,\quad t=r\sin\alpha.
\end{equation}
If we express the Einstein equations in terms of these coordinates and evaluate them at $A$, we find the regularity conditions
\begin{equation}
  r=0:\quad
 \frac{Q_{,\alpha}N_{,rr}^{\ \ 2}}{\sin\alpha\cos\alpha}=c,\quad
 N_{,rr}=\frac{c_1}{2}\sin^2\!\alpha+\frac{c^2}{8c_1}\cos^2\alpha,
\end{equation}
where $c:=\frac{8}{1-Q\p}$ and with a constant $c_1>0$. This can be used to show that $Q_{,tt}$ at $\Hp$ diverges as $1/\sin^2\!\theta$ at $A$ (i.e.\ as $\theta\to 0$) precisely in such a way that
\begin{equation}\label{eq:Acond}
 \Hp:\quad 
 \lim\limits_{\theta\to0}(\sin^2\!\theta\, Q_{,tt})N_{,\theta\theta}^{\ \ 2}
 =c\equiv\frac{8}{1-Q\p}.
\end{equation}
Similarly, one finds at $B$ (by introducing polar coordinates centred there)
\begin{equation}\label{eq:Bcond}
 \Hp:\quad 
 \lim\limits_{\theta\to\pi}(\sin^2\!\theta\, Q_{,tt})N_{,\theta\theta}^{\ \ 2}
 =-\frac{8}{1+Q\p}.
\end{equation}

From the point of view of an initial value problem for the Ernst equation, one can prescribe the initial Ernst potential $\E=f+\ii b$ at $t=0$. However, in order to guarantee that the solution will lead to a metric potential $Q$ with the correct boundary values at the axes $\Al$, $\Ar$ and the past horizon $\Hp$, it follows with the latter equations that the conditions
\begin{equation}
 b_{,\theta\theta}|_A=\frac{2Q\p}{Q\p-1},\quad
 b_{,\theta\theta}|_B=\frac{2Q\p}{Q\p+1}
\end{equation}
must be satisfied. (Note that, in the limit $a\p\to0$ and in terms of $x=\cos\theta$, these conditions become $b_{,x}|_A=-2$, $b_{,x}|_B=2$, which are precisely the conditions that have been used in \cite{BeyerHennig2014}, where an exact SGGTN solution has been derived.) This shows that the Ernst potential will be ill-behaved at $A$ or $B$ if $Q\p=\pm 1$. Since $\E$ is invariantly defined in terms of the Killing vectors, this indicates a physical problem, which justifies that we assume $Q\p\neq\pm1$ in our calculations, as already mentioned above.

The above relations, describing the behaviour of solutions near $A$ and $B$, will be used in the following to study the properties of the metric potentials after an appropriate rotation of the Killing basis.
\subsection{Rotation of the Killing basis}\label{sec:rotation}

The form \eqref{eq:metric2} of the metric is invariant under the coordinate transformation
\begin{equation}\label{eq:rot}
 \rho_1=\alpha\tilde\rho_1+\beta\tilde\rho_2,\quad
 \rho_2=\gamma\tilde\rho_1+\delta\tilde\rho_2,\quad
 \alpha, \beta, \gamma, \delta =\textrm{constant},
\end{equation}
but the potentials $v$, $Q$ and the constant $R_0$ change. Using the formulae in \cite{BeyerHennig2012} we obtain
\begin{eqnarray}
  \label{eq:rot1}\tilde R_0 &=|\alpha\delta-\beta\gamma| R_0,\\
 \label{eq:rot2}
  \ee^{\tilde v}&=\frac{(\alpha+\gamma\, Q)^2 \ee^v+\gamma^2 \ee^{-v}\sin^2\! t\,\sin^2\!\theta}{|\alpha\delta-\beta\gamma|},\\
 \label{eq:rot3}
  \tilde Q&=\frac{(\alpha+\gamma\,Q)(\beta+\delta\,Q)+\gamma\, \delta\, \ee^{-2v}\sin^2\! t\,\sin^2\!\theta}
  {(\alpha+\gamma\, Q)^2+\gamma^2 \ee^{-2v}\sin^2\! t\,\sin^2\!\theta}.
\end{eqnarray}
We intend to find a coordinate transformation of the above type for which $\partial_{\tilde\rho_1}=\partial_{\rho_1}-a\p\partial_{\rho_2}$, i.e.\ $\Hp$ is then generated by $\partial_{\tilde\rho_1}$. Since $\partial_{\tilde\rho_1}=\alpha\partial_{\rho_1}+\gamma\partial_{\rho_2}$, we have to choose
\begin{equation}
 \alpha=1,\quad \gamma=-a\p,
\end{equation}
so that Eq.~\eqref{eq:rot3} for $\tilde Q$ becomes
\begin{equation}
 \tilde Q 
 = \frac{(1-a\p Q)(\beta+\delta Q)\ee^{2v}-a\p \delta \sin^2\! t\,\sin^2\!\theta}
        {(1-a\p Q)^2\ee^{2v}+a\p^2\sin^2\!t\,\sin^2\!\theta}.
\end{equation}
The corresponding axes values are
\begin{equation}
 \tilde Q_1=\lim\limits_{\theta\to 0}\tilde Q = \frac{\beta+\delta}{1-a\p},\quad
 \tilde Q_2=\lim\limits_{\theta\to 0}\tilde Q = \frac{\beta-\delta}{1+a\p},
\end{equation}
and at $t=0$ we have
\begin{equation}
 \tilde Q(0,\theta)=-\frac{(\beta+\delta Q\p)Q_{,tt}N^2}{2a\p\sin^2\!\theta}-\frac{\delta}{a\p}.
\end{equation}
By virtue of \eqref{eq:Acond}, \eqref{eq:Bcond} it follows from the latter equations that
\begin{equation}
 \lim\limits_{\theta\to0}\tilde Q(0,\theta)=\tilde Q_1,\quad
 \lim\limits_{\theta\to\pi}\tilde Q(0,\theta)=\tilde Q_2,
\end{equation}
i.e., in contrast to $Q$, the transformed function $\tilde Q$ is continuous at $A$ and $B$.

A formula for $\ee^{\tilde u}$ follows from \eqref{eq:rot2},
\begin{equation}
 \ee^{\tilde u} 
 = \frac{1}{|\delta+\beta a\p|}\left[\left(\frac{1-a\p Q}{\sin t}\right)^2\ee^v+a\p^2\ee^{-v}\sin^2\!\theta\right],
\end{equation}
which shows that $\ee^{\tilde u}$ (unlike $\ee^u$) is regular at $\Hp$ as a consequence of the boundary conditions $Q=1/a\p$ and $Q_{,t}=0$ at $t=0$ (where the latter condition follows from the Einstein equations).

We still have the freedom to choose the constants $\beta$ and $\delta$, which can be used to obtain the same boundary values $\tilde Q_1=1$, $\tilde Q_2=-1$ as in our original coordinates. For that purpose, we need to choose
\begin{equation}
 \beta=-a\p,\quad \delta=1.
\end{equation}

Now we have almost reduced the problem with a general past Cauchy horizon to our earlier problem (the case with $a\p=0$, corresponding to a ``special'' past horizon), namely we have the same behaviour of regular functions $u$ and $Q$ near the past horizon. However, the boundary condition \eqref{eq:Mcond} for $M$ will change. Indeed, with the transformation formula for the constant $R_0$, 
\begin{equation}
 \tilde R_0=|\alpha\delta-\beta\gamma|R_0=|1-a\p^2|R_0,
\end{equation}
we obtain
\begin{eqnarray}
 \Al:\quad && \ee^{M+\tilde u}=\frac{(1-a\p)^2}{|1-a\p^2|}\ee^{M+u}
    =\left|\frac{1-a\p}{1+a\p}\right|R_0
    =\frac{\tilde R_0}{(1+a\p)^2},\label{eq:Mcond2a}\\
 \Ar:\quad && \ee^{M+\tilde u}=\frac{(1+a\p)^2}{|1-a\p^2|}\ee^{M+u}
    =\left|\frac{1+a\p}{1-a\p}\right|R_0
    =\frac{\tilde R_0}{(1-a\p)^2}\label{eq:Mcond2b}.
\end{eqnarray}
For $a\p=0$, this reduces to the earlier conditions \eqref{eq:Mcond}, but otherwise we have to use these modified boundary conditions.

Summarising our results so far, we see that we arrive almost at the same problem as in the case $a\p=0$. There are only two differences:
\begin{enumerate}
 \item The coodinates $(\tilde\rho_1,\tilde\rho_2)$ are defined in a different domain than $(\rho_1,\rho_2)$. In particular, the coordinate lines are in general (namely, for irrational $a\p$) not closed curves anymore. Hence, the past horizon, which is generated by $\partial_{\tilde\rho_1}$, has a null generator with non-closed orbits. 
 \item The boundary conditions for $M$ on the axes are different. We have to use \eqref{eq:Mcond2a}, \eqref{eq:Mcond2b} instead of \eqref{eq:Mcond}.
\end{enumerate}

Consequently, local existence of solutions to the $\tilde u$- and $\tilde Q$-equations can be shown with the Fuchsian methods as before without any change. However, the requirement of integrability of the $M$-equations will lead to a different initial constraint. This new condition can already be obtained as follows. Evaluating \eqref{eq:Mcond2a} at $A$ and \eqref{eq:Mcond2b} at $B$, together with evaluation of the boundary condition $\ee^{M-\tilde u}|_\Hp=\textrm{constant}$ (which follows from the Einstein equations) at $A$ and $B$, we obtain four equations involving function values at $A$ and $B$. These can be combined such that  $\ee^M|_A$, $\ee^M|_B$ and the constant are eliminated, leaving one condition involving only $\tilde u|_A$ and $\tilde u|_B$. This condition is
\begin{equation}\label{eq:ucond}
 |1+a\p|\ee^{\tilde u|_A}=|1-a\p|\ee^{\tilde u|_B}.
\end{equation}
For $a\p=0$, the latter equation implies $\tilde u|_A=\tilde u|_B$. In terms of the function $\tilde S_{**}$, which is related to $\tilde u$ at $\Hp$ via $R_0\ee^{\tilde u}=\ee^{\tilde S_{**}}$ [cf.~\eqref{eq:uexpan}], this leads to the periodicity condition $\tilde S_{**}(0)=\tilde S_{**}(\pi)$, which was required in that case. In the case of nonvanishing $a\p$, however, it follows from \eqref{eq:ucond} that this condition must be replaced by the ``jump condition''
\begin{equation}\label{eq:jumpcond}
 \tilde S_{**}(0)=\tilde S_{**}(\pi)+\ln\left|\frac{1-a\p}{1+a\p}\right|.
\end{equation}
Hence, for given $a\p$, we have to choose the asymptotic data function $\tilde S_{**}$ subject to this constraint. But we can also read \eqref{eq:jumpcond} differently. If we are not interested in a particular $a\p$, we can choose an arbitrary smooth function $\tilde S_{**}(\theta)$ and then determine the corresponding value of $a\p$ from 
\eqref{eq:jumpcond}, which tells us what the generator of the past horizon is. Note, however, that there are always two possible values for $a\p$, since  $a\p$ and $1/a\p$ lead to the same ``jump''. Physically, these correspond to the same situation, only expressed in different coordinates. Indeed, a solution with parameter $a\p$ can be transformed into one with parameter $1/a\p$ by an ``inversion'', i.e.\ an interchange of the Killing fields, corresponding to a coordinate transformation $\rho_1=\tilde\rho_2$, $\rho_2=\tilde\rho_1$. Under this transformation, according to \eqref{eq:rot3}, $Q$ is replaced by $Q/(Q^2+\ee^{-2v}\sin^2\! t\sin^2\!\theta)$ so that $Q\p\to1/Q\p$, which implies $a\p\to1/a\p$, cf.\ \eqref{eq:aQ}.

We can now generalise the definition of SGGTN solutions such that it incorporates the new situation with past Cauchy horizons ruled by non-closed null generators.
\begin{definition}[Extended definition of SGGTN solutions]\label{def2}
 SGGTN solutions are the class of all Gowdy-symmetric vacuum solutions with spatial topology $\Sth$ for which, in terms of the coordinates $t$, $\theta$, $\rho_1$, $\rho_2$ with line element \eqref{eq:metric}, the metric functions $u(t,\theta)$ and $Q(t,\theta)$ at $t=0$ are smooth functions on $\Sth$ in a Killing basis for which $Q$ has the boundary values $Q(0,0)=1$, $Q(0,\pi)=-1$.
\end{definition}
\begin{remark}\mbox{}

 \begin{enumerate}
  \item We have dropped the tildes at the coordinates and metric functions, but the definition refers to quantities with respect to the rotated Killing basis as discussed above. In particular, this implies that the coordinates $\rho_1$ and $\rho_2$ are no longer defined in the domains given in \eqref{eq:domains}, but in the region that is obtained after performing the transformations described in this subsection [cf.~Eq.~\eqref{eq:rot}] with the parameter $a\p$ that can be read off from the horizon data using \eqref{eq:ucond}.
  \item The only difference between this extended definition and the earlier definition \ref{def1} is that we have removed the periodicity condition. 
  \item Again, the class of solutions has two functional degrees of freedom, which can now be chosen without the periodicity requirement for $u$ (or, equivalently, $S_{**}$).
  \item Once again we have a large family of solutions with past Cauchy horizons, but there may be further smooth solutions with this property that are not contained in our class of SGGTN solutions.
 \end{enumerate}
\end{remark}

The above considerations show that \eqref{eq:jumpcond} is a \emph{necessary} condition for the existence of regular solutions to the $M$-equations. If we repeat the discussion of integrability of these equations as carried out in \cite{BeyerHennig2012}, thereby taking into account the slightly modified boundary conditions \eqref{eq:Mcond2a}, \eqref{eq:Mcond2b} for $M$, it then follows that \eqref{eq:jumpcond} is also a \emph{sufficient} condition. Hence we can indeed be sure that solutions exist locally. And once local existence is established, global existence and regularity of a future Cauchy horizon $\Hf$ at $t=\pi$ follow as in \cite{BeyerHennig2012}.

Summarizing, we have seen that SGGTN solutions exist for any smooth asymptotic data functions subject to the jump condition \eqref{eq:jumpcond}, which depends on the chosen generator of the past Cauchy horizon via the parameter $a\p$. 
The formulae for the Ernst potential at $\Al$, $\Ar$ and $\Hf$ in terms of the initial potential at $\Hp$ derived in \cite{BeyerHennig2012} can be carried over without change, as can the equations for $u$ and $Q$ at these boundaries (which will provide us with the potentials with respect to the rotated Killing basis, i.e.\ with $\tilde u$ and $\tilde Q$). Finally, $M$ at the boundaries can then be computed from \eqref{eq:Mcond2a}, \eqref{eq:Mcond2b} and from the formula $\ee^{M-\tilde u}|_\Hf=\textrm{constant}$ (which, as the above-mentioned similar equation at $\Hp$, can be derived from the first-order equations for $M$), where the constant follows from continuity at the intersection of $\Hf$ and the axes. 

The future Cauchy horizon $\Hf$ at $t=\pi$ is generated by a linear combination  of the Killing  vectors. In terms of the new coordinates, we can just copy the earlier results [cf. Eq.~\eqref{eq:af0}], which show that $\Hf$ is generated by 
$\partial_{\tilde\rho_1}-\tilde a_f\partial_{\tilde\rho_2}$ with
\begin{equation}
 \tilde a\f = \frac{8\Delta\tilde b}{16+\Delta\tilde b^2},\quad
 \Delta\tilde b:=\tilde b_B-\tilde b_A,
\end{equation}
where $\tilde b$ is the imaginary part of the Ernst potential $\tilde\E$ corresponding to the Killing vectors $\partial_{\tilde\rho_1}$ and $\partial_{\tilde\rho_2}$. The transformation equations for the metric potentials, together with the equations defining the Ernst potential, can be used to re-express the generator of $\Hf$ in terms of the original coordinates. We obtain that $\Hf$ is generated by
$\partial_{\rho_1}-a\f\partial_{\rho_2}$, where $a\f$ is determined by the initial Ernst potential and the constant $a\p$,
\begin{equation}\label{eq:af}
 a\f = a\p+8\,\frac{(1-a\p^2)\Delta b-4 a\p}{16+(1-a\p^2)\Delta b^2},\quad
 \Delta b:=b_B-b_A.
\end{equation}
Furthermore, just like in the case $a\p=0$, the solutions are regular with the exception of singular cases, which occur for $\Delta\tilde b=\pm 4$.

In the next subsection we construct a three-parametric family of exact solutions within this class of cosmological models.

\section{A family of exact solutions\label{sec:exact}}
\subsection{Construction of the solution}

We intend to solve an initial value problem for the Ernst equation in the coordinates $(\tilde\rho_1,\tilde\rho_2)$. For ease of notation, however, we will omit the tildes that have previously indicated quantities in these coordinates. 

We choose an initial Ernst potential $\E=f+\ii b$ of the form
\begin{equation}\label{eq:E0}
 t=0:\quad
 f=c_1(1-x^2)\left(1-\frac{x}{d}\right),\quad
 b=-x^2,
\end{equation}
where $x:=\cos\theta$, and $c_1>0$ and $d\in\R\setminus\{0\}$ are parameters. 
This should be compared to the initial potential for the family of solutions obtained in \cite{BeyerHennig2014}, where, as opposed to the present case, the real part was quadratic in $x$ and the imaginary part was cubic. Also note that \eqref{eq:E0} reduces to the initial Ernst potential of the Taub solution in the limit $d\to\infty$. 

The motivation for choosing the particular initial data \eqref{eq:E0} is that they are simple enough to allow for an exact solution of the integral equation that appears in ``Sibgatullin's integral method'', and simple enough to enable us to handle the rather long expressions that appear in intermediate steps of the calculations. On the other hand, \eqref{eq:E0} is sufficiently complex to provide a nontrivial example for a solution with a past horizon ruled by non-closed generators.

The above initial data have to satisfy certain conditions at the points $A$ ($t=\theta=0$) and $B$ ($t=0$, $\theta=\pi$). As a consequence of the definition of $f$ in terms of $u$ and $Q$, we have
\begin{equation}
  f|_A= f|_B=0,
\end{equation}
which is already satisfied by our above choice for $ f$.  
The $x$-derivatives of $b$ at $A$ and $B$ are related to the axis values of $Q$. Using that $Q_{1/2}=\pm 1$, we find
\begin{equation}
  b_{,x}|_A=-2,\quad  b_{,x}|_B=2,
\end{equation}
which are satisfied as well. Finally, the integrability condition \eqref{eq:ucond} for the $M$-equations translates into
\begin{equation}
 f_{,x}|_A=-\left|\frac{1+a\p}{1-a\p}\right| f_{,x}|_B.
\end{equation}
This fixes the above parameter $d$ in terms of $a\p$,
\begin{equation}
 d=\left\{\begin{array}{ll}\displaystyle
             -\frac{1}{a\p}, & |a\p|<1\\
             -a\p, & |a\p|>1
            \end{array}\right. .
\end{equation}
Hence, we always have $|d|>1$.

The initial value problem for the Ernst equation with initial Ernst potential \eqref{eq:E0} can be solved with ``Sibgatullin's integral method'' \cite{Sibgatullin1984, MankoSibgatullin1993}.
Since this procedure is described at length in \cite{BeyerHennig2014}, we refer to that paper for details and give here only the solution to the present initial value problem. The result is the following Ernst potential,
\begin{eqnarray}
 \fl \E & =\Bigg[16 d^4 (1-y)^2 (d-xy)-c_1^4 (1-x^2)^2 (1+y)^6 (d-x y)\nonumber\\*
 \fl & \quad -8 c_1^2 d^2 (1+y)^2 \Big(2 d^3+d^2 x (8-6 y)-d [(1-y) (7+3 y)-3 x^2 (1-4y+y^2)]\nonumber\\*
 \fl & \quad -x [8-7y-4y^2+3 y^3-x^2 (8-3y-4y^2+y^3 )]\Big)\nonumber\\*
 \fl &\quad  +8 \ii c_1 d^3 (1-y^2) \Big(4 d^2+8 d x (1-y)-(1-y) (5+y)+x^2(9-4y+3 y^2)\Big)\nonumber\\*
 \fl & \quad +2 \ii c_1^3 d (1-x^2) (1+y)^4 
     \Big(1+8 d^2+8 d x (1-2 y)-4 y+3 y^2+x^2(11-12y+5y^2)\Big)\nonumber\\*
 \fl &\quad -r_d \Big(4 d^2 (1-y)+4 \ii c_1 d [d+x (2-y)] (1+y)+c_1^2 (1-x^2) (1+y)^3\Big)^2\Bigg]\nonumber\\*
 \fl & \quad /
     \Big[8 c_1 d (1+y) \Big(4 d^2 (1-y)-c_1^2 (1-x^2)(1+y)^3
     +4 \ii c_1 d (1+y) (d-x y)\Big)\Big],
\end{eqnarray}
where $x=\cos\theta$, $y=\cos t$ and
\begin{equation}
 r_d:=\mathrm{sgn}(d)\sqrt{(d-xy)^2-(1-x^2)(1-y^2)}.
\end{equation}

In a next step we construct the corresponding metric potentials $u$, $Q$ and $M$. From $f=\Re(\E)$ and the Ernst auxiliary quantity $a$ (the procedure for computing $a$ is explained in \cite{BeyerHennig2014}), we obtain the metric potentials $u$ and $Q$ via
\begin{equation}\label{eq:uQ}
 \ee^u=\frac{f a^2}{\sin^2\! t}+\frac{\sin^2\!\theta}{f},\quad
 Q=\frac{f^2 a}{f^2 a^2 + \sin^2\! t\sin^2\!\theta}.
\end{equation}
Afterwards, the function $M$ is obtained via line integration from $u$ and $Q$ using the first-order Einstein equations for $M$. The results are
\begin{eqnarray}\nopagebreak
 \fl \ee^u &=& 8 c_1 d
       \frac{ 4d^2(d+r_d-xy)+c_1^2(1-x^2)^2(1+y)^4/(d+r_d-xy)}
       {\big[4d^2(1-y)+c_1^2(1-x^2)(1+y)^3\big]^2+\big[4c_1 d(1+y) r_d\big]^2},\\
 \fl Q &=& x+\frac12(d-r_d-xy) \equiv x+\frac{(1-x^2)(1-y^2)}{2(d+r_d-xy)},\\
 \fl \ee^M &=&
       \frac{c}{r_d}\left[4d^2(r_d + x - dy)^2 + 
       c_1^2 (1 + y)^2 \Big(r_d^2 + (d x - y) (1 + y) + (d + x) r_d\Big)^2\right].
\end{eqnarray}
The integration constant $c$ in the expression for $M$ is fixed by the axis boundary conditions \eqref{eq:Mcond2a}, \eqref{eq:Mcond2b}. (One of these two conditions already fixes $c$. However, the other condition is then automatically satisfied as well, since we have chosen our initial data subject to the jump condition such that the $M$-equations are integrable.) We obtain
\begin{equation}
 c=\frac{R_0}{16 c_1 d(1+a\p)^2(1+d)^2}
  =\left\{\begin{array}{ll}\displaystyle
           \frac{dR_0}{16 c_1(1-d^2)^2},& |a\p|<1\\
            \displaystyle
           \frac{R_0}{16 c_1 d(1-d^2)^2}, & |a\p|>1
          \end{array}\right. .
\end{equation}
Note that $0<\ee^u<\infty$ and $0<\ee^M<\infty$ holds within the entire Gowdy square $t,\theta\in[0,\pi]$, i.e.\ $x,y\in[-1,1]$, and the solutions are regular for all choices of the parameters. Indeed, the singular cases (with curvature singularities at the boundary $t=\pi$) would only occur for initial data with $b_B-b_A=\pm4$. Here, however, we have $b_B-b_A=0$.
\subsection{Extensions}

In this subsection, we show how the solution can be extended through the Cauchy horizons $\Hp$ and $\Hf$, and we study properties of the regions beyond these horizons.

In a first step, we rewrite the metric \eqref{eq:metric} in terms of $x=\cos\theta$ and $y=\cos t$, which is reasonable since the metric functions are most naturally given as functions of $x$ and $y$ as well, but which is also useful for the purpose of constructing extensions through the Cauchy horizons (as $y$ turns out to be a ``better'' time coordinate than $t$).  We obtain
\begin{equation}\fl\label{eq:metric3}
 g_{ab}=\ee^M\left(-\frac{\dd y^2}{1-y^2}+\frac{\dd x^2}{1-x^2}\right)
        +R_0\left[(1-y^2)\ee^u (\dd\rho_1+Q \dd\rho_2)^2+(1-x^2)\ee^{-u} \dd\rho_2^2\right].
\end{equation}
This form of the metric is clearly singular at $x=\pm1$ (corresponding to coordinate singularities at the symmetry axes $\Al$, $\Ar$, which could be removed by locally introducing Cartesian coordinates after going back to the original Killing basis, as described in the appendix of \cite{BeyerHennig2014}) and at $y=\pm 1$ (corresponding to coordinate singularities at the future and past Cauchy horizons). In order to remove the latter singularities, we replace $\rho_1$ by a new coordinate $\rho_1'$, defined by
\begin{equation}\label{eq:trans1}
 \rho_1' = \rho_1-\kappa\ln|1-y|-\kappa_1\ln|1+y|
\end{equation}
with constants $\kappa$ and $\kappa_1$. 
Note that the absolute values in \eqref{eq:trans1} would not be necessary at this stage, since $1\pm y$ is positive in the interior of the Gowdy square. However, we intend to use the same transformation later in the regions $|y|>1$, where the absolute values will ensure that we have a real transformation.
The new metric is
\begin{eqnarray}
 \fl
 g_{ab} &=& \ee^M\left(-\frac{\dd y^2}{1-y^2}+\frac{\dd x^2}{1-x^2}\right)\nonumber\\ \fl
       && +R_0\Bigg[(1-y^2)\ee^u\left(\dd\rho_1'+Q\dd\rho_2
            -\frac{\kappa\dd y}{1-y}+\frac{\kappa_1\dd y}{1+y}\right)^2
         +(1-x^2)\ee^{-u}\dd\rho_2^2\Bigg].
\end{eqnarray}
The components
\begin{equation}
 g_{y\rho_1'}=R_0\ee^u[\kappa_1(1-y)-\kappa(1+y)]\quad\textrm{and}\quad
 g_{y\rho_2}=Qg_{y\rho_1'}
\end{equation}
are regular at $y=\pm 1$, and 
\begin{equation}
 g_{yy}=\frac{R_0\ee^u[\kappa_1(1-y)-\kappa(1+y)]^2-\ee^M}{1-y^2}
\end{equation}
is regular as well, provided we choose
\begin{equation}
 \kappa^2=\lim\limits_{y\to 1}\frac{\ee^{M-u}}{4R_0}=\frac{4cc_1^3(1-d^2)^2}{dR_0}\quad\textrm{and}\quad
 \kappa_1^2=\lim\limits_{y\to-1}\frac{\ee^{M-u}}{4R_0}=\frac{4cd^3}{c_1R_0}.
\end{equation}
With the above value for $c$ this leads to
\begin{equation}
 \kappa=\pm\left\{\begin{array}{ll}
                   \frac{c_1}{2}, & |a\p|<1\\
                   \frac{c_1}{2d},& |a\p|>1
                  \end{array}\right.
 \quad\textrm{and}\quad
 \kappa_1=\pm\left\{\begin{array}{ll}
                   \frac{d^2}{2c_1(1-d^2)}, & |a\p|<1\\
                   \frac{d}{2c_1(1-d^2)},& |a\p|>1
                  \end{array}\right. .
\end{equation}
The signs for $\kappa$ and $\kappa_1$ can be chosen independently, which leaves us with four possible coordinate transformations and, therefore, with four extensions. This is exactly in line with the four ``standard extensions'' of the Taub-NUT solution (and with the four extensions of the solution family studied in \cite{BeyerHennig2014}). The metric \eqref{eq:metric3} is now valid for all values of $y$, where we choose the analytic extensions of the metric potentials to $y\in\R$.

The properties $0<\ee^M<\infty$ and $0<\ee^u<\infty$ also hold almost everywhere in the extended regions, and because of $\det(g_{ab})=-R_0^2\,\ee^{2M}$, the new metric is invertible even at $y=\pm 1$. The only exceptions are the two half-lines 
\begin{equation}
 \Hps:x=\mathrm{sgn}(d),\ y>|d|\quad\textrm{and}\quad
 \Hfs:x=-\mathrm{sgn}(d),\ y<-|d|,
\end{equation}
where $\ee^u$ vanishes, and, in particular, the endpoints of these half-lines,
\begin{equation}
 \mathcal S\p: x=\mathrm{sgn}(d),\ y=|d|\quad\textrm{and}\quad
 \mathcal S\f: x=-\mathrm{sgn}(d),\ y=-|d|,
\end{equation}
where $r_d=0$ holds, such that $\ee^M$ diverges at these two points. 
Moreover, the function $Q$, which otherwise has the axis boundary values $\pm1$ at $x=\pm1$, takes on the values $\mathrm{sgn}(d)(1-y)+d$ on $\Hps$ and $\Hfs$.
We will see that this behaviour corresponds to coordinate singularities that can be removed. Henceforth, we will always assume that $d>0$, in order to avoid the term $\mathrm{sgn}(d)$. With minor modifications the same considerations can be performed for negative $d$.

Once we have extended the solution to $|y|>1$, we can restore the form \eqref{eq:metric3} of the metric by undoing the coordinate transformation \eqref{eq:trans1}. This metric is then only valid separately in the regions $y<-1$, $-1<y<1$ and $y>1$, but not in the entire range $y\in\R$. Nevertheless, it is still preferable to use this simple form of the metric and to accept the limited range of validity.

In order to illustrate the properties of the regions that we have already discovered and further regions that will be presented shortly, we follow the journey of an observer who starts inside the Gowdy square and moves into the region $y>1$.

If we assume that our original time coordinate $t$ increases with time, then $y=\cos t$ would \emph{decrease}. Hence, a journey into the region $y>1$ would actually be a journey into the past of the observer. However, since it is convenient to think of an upward world line in a spacetime diagram as corresponding to a future-directed motion, we will just pretend that we look at the future of our observer. Instead, we could also follow an observer who travels into the region $y<-1$, but then additional minus signs would appear in some of the following transformations, which we avoid by restricting to the region $y>1$.

\begin{figure}
 \centering
 \includegraphics[scale=0.8]{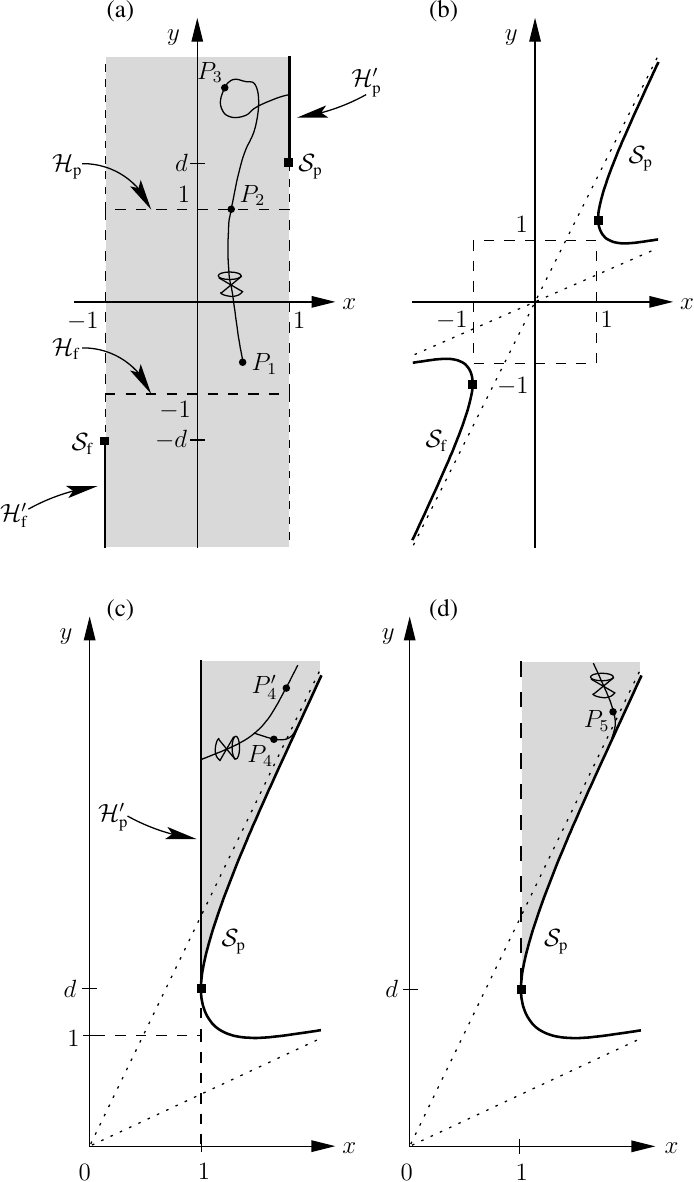}
 \caption{Illustration of the causal structure of the solution. Details are given in the text.\label{fig:Ext}}
\end{figure}

Our observer starts at point $P_1$ in Fig.~\ref{fig:Ext}a and moves in positive $y$-direction. A light cone is sketched in Fig.~\ref{fig:Ext}a to illustrate that $y$ plays the role of a time coordinate within the Gowdy square. The above regularisation of the metric has shown that the observer can reach the point $P_2$ on the Cauchy horizon $\Hp$ without problems, and he will arrive there after a finite proper time. In the extended region $y>1$ the coefficient of $\dd y^2$ in \eqref{eq:metric3} becomes positive, thus $y$ is no longer a time coordinate. Instead, $\rho_1$ becomes the new time coordinate. Therefore, our observer can move to arbitrary values of $x$ and $y$ in this region. (For example, he could do a loop as shown near point $P_3$.) Furthermore, it is not possible to sketch a light cone in this domain, since the time direction is not contained in the $x$-$y$-plane anymore.  Most importantly, because of the timelike nature of $\partial_{\rho_1}$ and the periodicity of the $\rho$-coordinates, there are closed causal curves in the extended regions --- precisely as observed for the NUT regions of the Taub-NUT solution. The figure also shows the singular boundary $\Hps$ and the singular point $\Sp$ (as well as $\Hfs$ and $\Sf$).

An analysis of the behaviour of null geodesics near $\Hfs$ suggests a coordinate transformation 
\begin{equation}
 \rho_1'=\rho_1-\mu\ln|1-x|,
\end{equation}
which should be compared to the very similar transformation \eqref{eq:trans1}. It turns out that the transformed metric is regular at $\Hps$, provided we choose the constant $\mu$ as
\begin{equation}
 \mu^2=\lim\limits_{\scriptsize\begin{array}{c}
                    x\to 1\\ y>d
                   \end{array}}
        \frac{\ee^M}{2R_0(y^2-1)\frac{\ee^u}{1-x}},
\end{equation}
which leads to the two possibilities (corresponding to two possible extensions)
\begin{equation}
 \mu=\pm\left\{\begin{array}{ll}
                \frac{d^2}{4c_1(1+d)}, & |a\p|<1\\
                \frac{d}{4c_1(1+d)}, & |a\p|>1
               \end{array}.\right.
\end{equation}
Note that the point $\Sp$, where $r_d=0$, is still singular in the new coordinates. However, we will shortly see that $\Sp$ is just a special point on an entire singular curve, whose singular nature can be ``cured'' with another coordinate transformation.

We observe that the Killing vectors $\partial_{\rho_1'}=\partial_{\rho_1}$ and $\partial_{\rho_2}$ satisfy
\begin{equation}
 \Hps:\quad g(\partial_{\rho_1'},\partial_{\rho_1'})=0,\quad
            g(\partial_{\rho_2},\partial_{\rho_2})=\frac{d(y-1)(y-d)}{c_1(1+y)}\neq 0,
\end{equation}
i.e.\ they are not linearly dependent on $\Hps$ in contrast to the ``regular'' parts of the axes. This shows that $\Hps$ is actually not a part of a symmetry axis. Instead, it corresponds to another Cauchy horizon, namely, a Cauchy horizon for the region $x>1$, which becomes accessible in terms of our new coordinates (this time by choosing the analytic continuations of the metric potentials to values $x>1$). Consequently, our observer could choose to travel through $\Hps$. Note that like $\Hp$, this Cauchy horizon is also generated by $\partial_{\rho_1}$. 

Once we have entered the region $x>1$, we again restore the coordinate $\rho_1$ and our earlier form of the metric by undoing the previous coordinate transformation, in order to keep the expressions simple.

In the regions with $|x|\le1$, the argument of the square root $r_d$ is always positive --- with the exception of the points $\Sp$ and $\Sf$, where $r_d=0$ holds. What happens beyond $\Hps$? The equation
\begin{equation}
 r_d^2 \equiv x^2+y^2-2dxy+d^2-1=0
\end{equation}
describes a hyperbola whose two branches asymptotically approach the lines $y=(d\pm\sqrt{d^2-1})x$. We denote the branches by the same symbols $\Sp$ and $\Sf$ (see Fig.~\ref{fig:Ext}b) as the earlier discussed singular points, in order to illustrate that these points are just special points on the curves at which $r_d$ vanishes. The hyperbola is always outside the region $|x|<1$ and touches the lines $x=\pm1$ precisely at the earlier noted singular points. Inside the hyperbola branches, $r_d^2$ is negative, so that the solution has no chance to be extended into that region, where the metric coefficients would be complex.

Our observer, who has now entered the region $x>1$, can travel through the shaded region in Fig.~\ref{fig:Ext}c. Here, the function $\ee^u$ is negative and also the coefficient of $\dd x^2$ in \eqref{eq:metric3} changes its sign. As a consequence, $\rho_1$ again becomes a spacelike coordinate, whereas $x$ is the new time coordinate. This is illustrated by the light cone shown in Fig.~\ref{fig:Ext}c.
Moreover, both Killing vectors are again spacelike as in the Gowdy square, where we started our journey.
The observer could now decide to travel to point $P_4'$ and further along a worldline that always avoids the hyperbola, i.e.\ he could remain forever in the shaded region. But he could also travel to point $P_4$ and further towards the hyperbola. What happens there?

It turns out that the divergence of $g_{xx}$ and $g_{yy}$ (as a consequence of a diverging $\ee^M$) is due to a coordinate singularity. Roughly speaking, we have used the square of a ``regular'' coordinate near the hyperbola instead of the regular coordinate itself. And besides introducing diverging metric components, this also means that we only cover the region where the regular coordinate is positive (or negative), but not both regions at the same time. This can be made more explicit with the following considerations.

Based on null geodesics in the $x$-$y$-plane, which can be obtained from the equation $\dd x^2/(1-x^2)-\dd y^2/(1-y^2)=0$, we can introduce null coordinates $\alpha$ and $\beta$ as follows,
\begin{eqnarray}
 x&=&\sqrt{\frac12\left(\alpha\beta+1-\sqrt{(\alpha^2-1)(\beta^2-1)}\right)},\\
 y&=&\sqrt{\frac12\left(\alpha\beta+1+\sqrt{(\alpha^2-1)(\beta^2-1)}\right)}.
\end{eqnarray}
They are defined for $\alpha\ge d$, $\beta\ge\alpha$. The upper half of $\Sp$ is given by $\alpha=d$ and $\beta\ge d$. In terms of these coordinates, $r_d$ takes the simple form
\begin{equation}
 r_d=\sqrt{(\alpha-d)(\beta-d)}.
\end{equation}
Moreover, the $x$-$y$-part of the metric becomes
\begin{equation}
 \ee^M\left(\frac{\dd x^2}{1-x^2}-\frac{\dd y^2}{1-y^2}\right)
 \propto\frac{\dd\alpha\,\dd\beta}{\sqrt{(\alpha-d)(\beta-d)}}.
\end{equation}
If we perform the additional coordinate transformation 
\begin{equation}
 \tilde\alpha=\sqrt{\alpha-d},\quad
 \tilde\beta=\sqrt{\beta-d},
\end{equation}
then we find
\begin{equation}
 \frac{\dd\alpha\,\dd\beta}{\sqrt{(\alpha-d)(\beta-d)}}
 \propto \dd\tilde\alpha\,\dd\tilde\beta,
\end{equation}
which is not singular anymore. Hence, $\tilde\alpha$ and $\tilde\beta$ can be used as the desired ``regular'' coordinates. Also note that 
\begin{equation}\label{eq:rd}
r_d=\tilde\alpha \tilde\beta. 
\end{equation}

The region above the hyperbola branch $\Sp$ corresponds to coordinate values $\tilde\alpha>0$. However, since nothing special happens at $\tilde\alpha=0$ in terms of the new coordinates, we may extend the solution to $\tilde\alpha\le 0$. As usual, we can afterwards undo the coordinate transformation to recover our original coordinates; here $x$ and $y$. The region $\tilde\alpha<0$ ``beyond'' $\Sp$ corresponds to the same coordinate values of $x$ and $y$ as the shaded region in Fig.~\ref{fig:Ext}c. However, because of \eqref{eq:rd}, the square root $r_d$ has to be taken with the opposite sign in the new region. Consequently, the values of the metric potentials change as well. Moreover, we still have that $\ee^u<0$, but also $\ee^M$, which was positive for $\tilde\alpha>0$, is now negative. This leads to a sign change of the metric components $g_{xx}$ and $g_{yy}$, such that $x$ is now a spatial coordinate and $y$ is again the time coordinate. This is illustrated in Fig.~\ref{fig:Ext}d, where the light cone has a different orientation compared to Fig.~\ref{fig:Ext}c. As a consequence, our observer now necessarily travels in the positive $y$-direction and cannot return to the hyperbola.

Note that the half-line $x=1$, $y>d$ (which corresponded to a Cauchy horizon in the previous domain) is a regular axis in the new domain. In particular, $\ee^u\neq0$ there and $Q$ has again the axis value $Q=1$. Also, both Killing vectors have the same positive norm there.

An interesting difference between the two regions with two \emph{spacelike} Killing vectors we have looked at, namely the Gowdy square and the region behind $\Hps$ (which consists in the two subregions shown in Figs.~\ref{fig:Ext}c, d) is the topology. Whereas the spatial topology in the Gowdy square is $\Sth$, a spacelike hypersurface beyond $\Hps$ of the form $F(x,y)=0$ has the topology $\R^2\times\So$. Thereby, $\So$ corresponds to one of the two symmetries, so that nothing depends on the position on the circle. Moreover, the $\R^2$-factor is rotationally symmetric, i.e.\ one could introduce polar coordinates such that the angular direction corresponds to the second symmetry and everything depends only on the radial coordinate. The origin of the polar coordinates is at the axis $x=1$ in Fig.~\ref{fig:Ext}d. Furthermore, the radial coordinate can either be unbounded (if we choose a spacelike hypersurface that extends up to $y=\infty$ in the region in Fig.~\ref{fig:Ext}c) or bounded (if the slice hits the horizon $\Hps$). Hence, unlike the spatially compact Gowdy square, the region beyond $\Hps$ is spatially unbounded and therefore not a cosmological model in the usual sense.

\section{Discussion\label{sec:discuss}}

We have generalized the local and global existence results for smooth Gowdy-symmetric generalized Taub-NUT solutions \cite{BeyerHennig2012} to the situation where the past Cauchy horizon at $t=0$ has non-closed null generators. We find that these solutions have again two functional degrees of freedom (asymptotic data), which must be chosen subject to a jump condition [cf.~\eqref{eq:jumpcond}] instead of the earlier required periodicity condition [cf.~\eqref{eq:percon}]. The size of the jump depends on the generator of the past horizon, expressed in the form of a parameter $a\p$. Alternatively, one can even choose arbitrary smooth asymptotic data functions, which then determine the parameter $a\p$ and therefore the generator of the past horizon. The application of the earlier results for past horizons with closed orbits to this new situation was possible with a suitable rotation of the Killing basis, which reduced this more general situation (with $a\p\neq0$) almost to the problem with $a\p=0$. Only a boundary condition for one of the metric potentials changed. We can also carry over the explicit construction of the metric at $t=\pi$ in terms of the data at $t=0$, which shows that the solutions develop a second Cauchy horizon there. As before, the only exceptions are singular cases in which curvature singularities form at the boundary $t=\pi$. Whether this is the case can be read off from the data at the past horizon. 

Furthermore, we have constructed a three-parametric family of exact solutions within our generalized class of cosmological models. It contains the spatially homogeneous Taub solution as the special case $a\p=0$. Otherwise, for $a\p\neq0$, we obtain explicit examples for inhomogeneous cosmological models. Moreover, we have constructed several extensions of the solutions through the Cauchy horizons, which opened up regions with closed causal curves. An interesting observation was the existence of further Cauchy horizons in these extensions, which allow access into regions where both Killing vectors are again spacelike. This shows that the causal structure of this family of solutions is more complex than the structure of the family of solutions presented in \cite{BeyerHennig2014}.

\section*{Acknowledgments}
 I would like to thank Florian Beyer and Gerrard Liddell for many valuable discussions and Chris Stevens for commenting on the manuscript. This work was supported by the Marsden Fund Council from Government funding, administered by the Royal Society of New Zealand.
 



\section*{References}

\begin{thebibliography}{10}

\bibitem{Ames2013a}
 Ames, E., Beyer, F., Isenberg, J., LeFloch, P.\ G.,
 \emph{Quasilinear hyperbolic Fuchsian systems and AVTD behaviour in $T^2$-symmetric vacuum spacetimes}, 
 Ann.\ Henri Poincar\'e {\bf 14}, 1445 (2013)

\bibitem{Ames2013b}
 Ames, E., Beyer, F., Isenberg, J., LeFloch, P.\ G.,
 \emph{Quasilinear symmetric hyperbolic Fuchsian systems in several space dimensions},
 Contemp.\ Math. {\bf 591}, 25 (2013)

\bibitem{BeyerHennig2012}
 Beyer, F.\ and Hennig, J., 
 \emph{Smooth Gowdy-symmetric generalized Taub-NUT solutions},
 Class.\ Quantum Grav.\ {\bf 29}, 245017 (2012)

\bibitem{BeyerHennig2014}
 Beyer, F.\ and Hennig, J.,
 \emph{An exact smooth Gowdy-symmetric generalized Taub-NUT solution},
 Class.\ Quantum Grav.\ {\bf 31}, 095010 (2014)

\bibitem{Chrusciel1990}
 Chru\'{s}ciel, P.\ T.,
 \emph{On space-times with $U(1)\times U(1)$ symmetric compact Cauchy
  surfaces},
  Ann.\ Phys.\ {\bf 202}, 100 (1990)

\bibitem{ChruscielIsenberg1993}
 Chru\'{s}ciel, P.~T.\ and Isenberg, J.,
 \emph{Nonisometric vacuum extensions of vacuum maximal globally hyperbolic
  spacetimes},
 Phys.\ Rev.\ D {\bf48}, 1616 (1993)

\bibitem{FriedrichRaczWald1999}
 Friedrich, H., R\'acz, I.\ and Wald, R.,
 \emph{On the rigidity theorem for spacetimes with a stationary event horizon or a compact Cauchy horizon},
 Commun.\ Math.\ Phys.\ {\bf 204}, 691 (1999)

\bibitem{MankoSibgatullin1993}
 Manko, V.\ S.\  and Sibgatullin, N.\ R.,
 \emph{Construction of exact solutions of the Einstein-Maxwell equations  corresponding to a given behaviour of the Ernst potentials on the symmetry axis},
 Class.\ Quantum Grav.\ {\bf 19}, 1383 (1993)

\bibitem{Misner1963}
 Misner, C.~W.,
 \emph{The flatter regions of Newman, Unti, and Tamburino's generalized Schwarzschild space},
 J.\ Math.\ Phys.\ {\bf 4}, 924 (1963)

\bibitem{MisnerTaub1969}
 Misner, C.~W. and Taub, A.~H.,
 \emph{A singularity-free empty universe},
 Sov. Phys. JETP {\bf 28}, 122 (1969)

\bibitem{Moncrief1984}
 V.\ Moncrief,
 \emph{The space of (generalized) Taub-NUT spacetimes},
 J.\ Geom.\ Phys.\ {\bf  1}, 107 (1984)

\bibitem{MoncriefIsenberg1983}
 Moncrief, V.\ and Isenberg, J.,
 \emph{Symmetries of cosmological Cauchy horizons},
 Commun.\ Math.\ Phys.\ {\bf 89}, 387 (1983)

\bibitem{Racz2000}
 R\'acz, I.,
 \emph{On further generalization of the rigidity theorem for spacetimes with a stationary event horizon or a compact Cauchy horizon},
 Class.\ Quantum Grav.\ {\bf 17}, 153 (2000)

\bibitem{Sibgatullin1984}
 Sibgatullin, N.\ R.,
 \emph{Oscillations and waves},
 Springer (Berlin, 1984)

\bibitem{Taub1951}
 Taub, A.~H.,
 \emph{Empty space-times admitting a three parameter group of motions},
 Ann.\ Math. {\bf 53}, 472 (1951)

\end{thebibliography}
\end{document}